\documentclass[pre,showpacs,nofootinbib,twocolumn,floatfix]{revtex4}

\usepackage{amsmath}
\usepackage{amssymb}
\usepackage{epsfig}
\usepackage{dcolumn}% Align table columns on decimal point
\usepackage{bm}     % bold math

\newcommand{\rB}{{\operatorname{B}}}
\newcommand{\VECr}{\boldsymbol{r}}
\newcommand{\VECx}{\boldsymbol{x}}
\newcommand{\re}{{\rm {e}}}

\newcommand{\rDHH}{{\rm {DHH}}}
\newcommand{\rDHHC}{{\rm {DHHC}}}
\newcommand{\rDHHCz}{{\rm DHHC}$^{(0)}$}\newcommand{\rOCP}{{\rm {OCP}}}
\newcommand{\infd}{{\mbox{d}}}

\newcommand{\rp}{{\rm {p}}}
\newcommand{\D}{\displaystyle}
\newcommand{\rPB}{{\rm {PB}}}
\newcommand{\rMC}{{\rm {MC}}}
\newcommand{\corr}{{\operatorname{corr}}}
\newcommand{\rel}{{\operatorname{el}}}

\newcommand{\CALH}{{\mathcal H}}
\newcommand{\CALT}{{\mathcal T}}
\newcommand{\CALV}{{\mathcal V}}
\newcommand{\CALZ}{{\mathcal Z}}

\newcommand{\nhat}{\widehat{n}}

\begin{document}

\preprint{PREPRINT}

\date{\today} 

\title{Screening of Spherical Colloids beyond Mean Field ---\\%
       A Local Density Functional Approach} 

\author{Marcia C.\ Barbosa}\email{barbosa@if.ufrgs.br}\affiliation{Instituto de
  F\'{\i}sica, UFRGS, 91501-970, Porto Alegre, RS, Brazil} 
\author{Markus Deserno}\email{deserno@mpip-mainz.mpg.de}\affiliation{Max-Planck-Institut
  f\"ur Polymerforschung, Ackermannweg 10, 55128 Mainz, Germany} 
\author{Ren\'e Messina}\email{messina@thphy.uni-duesseldorf.de} \affiliation{Institut f\"ur
  theoretische Physik II, Heinrich-Heine-Universit\"at D\"usseldorf,
  Universit\"atsstr.\ 1, D-40225 D\"usseldorf, Germany}
\author{Christian Holm}\email{holm@mpip-mainz.mpg.de}\affiliation{Max-Planck-Institut f\"ur
  Polymerforschung, Ackermannweg 10, 55128 Mainz, Germany} 

\begin{abstract}
  We study the counterion distribution around a spherical macroion and
  its osmotic pressure in the framework of the recently developed
  Debye-H\"uckel-Hole-Cavity (DHHC) theory.  This is a local density
  functional approach which incorporates correlations into
  Poisson-Boltzmann theory by adding a free energy correction based on
  the One Component Plasma.  We compare the predictions for ion
  distribution and osmotic pressure obtained by the full theory and by
  its zero temperature limit with Monte Carlo simulations.  They agree
  excellently for weakly developed correlations and give the correct
  trend for stronger ones.  In all investigated cases the DHHC theory
  and its computationally simpler zero temperature limit yield better
  results than the Poisson-Boltzmann theory.
\end{abstract}


\maketitle

%%%%%%%%%%%%%%%%%%%%%%%%%%%%%%%%%%%%%%%%%%%%%%%%%%%%%%%%%%%%%%%%%%%%%%%%%%%%%%

\section{Introduction}

The screening of charged macromolecules in an electrolyte solution is
a long standing problem which has prompted many attempts aiming at a
theoretical explanation.  In their pioneering work Gouy \cite{gouy10a} and
Chapman \cite{chapman13a} used what is now referred to as
Poisson-Boltzmann (PB) theory as the basis for a mean field treatment of
the electrical double layer.  This approach found its culmination
about thirty years later in the famous DLVO theory of charged colloids
\cite{derjaguin41a,verwey48a}.  The major flaw of these mean field
approaches is their neglect of correlations between the ions.  The
first attempt to work out such correlations for \emph{homogeneous}
electrolytes are due to Debye and H\"uckel \cite{debye23a}, whose work
remarkably (and at first glance confusingly) is also based on
(linearized) Poisson-Boltzmann theory.  In the \emph{inhomogeneous}
case integral equation theories
\cite{belloni86a,hansen90a,groot91b,lozada92b} and recently field
theories \cite{netz99a} have become very popular in calculating
correlation corrections to mean field double layers.  However, in
order to make progress and calculate physical quantities,
approximations have to be made which, unfortunately, instead of
clarifying the physics sometimes tend to obscure it.  Moreover, since
in some of these methods, the free energy is not defined in a unique
way, it becomes impossible to determine the specific role played by
each source of correlations in the system.

It would therefore be desirable to have a theoretical framework which
retains the simplicity of the early attempts, but also accommodates
correlation effects.  This is the case for density functional
theories.  It is possible to rigorously rewrite the partition function
of, say, a system of charged colloids, as a density functional
\cite{loewen93a}, in which the contribution beyond mean field is seen
to be expressible as an additive correlation correction to the free
energy density, whose functional form is of course unknown and for
which one has to make a reasonable ansatz.  The spirit is very similar
to the fundamental problem of integral equations, where one also has
to make an educated guess (namely, the closure relation), but in the
functional case the ansatz involves a free energy density rather than
a relation between two- and three-point functions.  It thus relies on
a different kind of intuition and thus permits complementary insight.

One suggestion for such a functional correction has been made by
Nordholm \cite{nordholm84a}.  It relies on a Debye-H\"uckel treatment
of the One Component Plasma (OCP) \cite{salpeter58a,abe59a,baus80a},
in which the short-distance failure of linearization is cleverly
overcome by postulating a correlation hole.  Since beyond a certain
density the resulting OCP free energy density is a concave function of
density, this favors the development of inhomogeneities.  In the pure
OCP these are balanced by the homogeneously charged background.
However, if one uses the OCP free energy density as a correlation
correction to the mean field functional describing the double layer at
a charged surface, one has all the charge opposite to the counterions
located on that surface, rather than homogeneously distributed as a
stabilizing background.  The consequence is that the double layer
becomes unstable and all ions collapse onto the surface, an effect
which has been termed ``structuring catastrophe''
\cite{groot91a,penfold90a}.

To circumvent this instability without losing the physical
transparency of a local functional, we recently proposed the
Debye-H\"uckel-Hole-Cavity (DHHC) theory \cite{barbosa00a}, in which we
suggested a convex correlation functional. This was achieved by
excluding the homogeneous background from a region of radius $a$
around the central ion during the Debye charging process.  For
counterions with size we identified $a$ tentatively as the ion
diameter.  We then applied our theory to the screening of a charged
rod by its counterions.  Comparisons of the ionic charge distribution
obtained showed a very good agreement with the simulations for both
monovalent and trivalent counterions.

In this paper we test our theory for a different geometry: charged
spherical colloids with point-like counterions.  In general, colloidal
systems exhibit a rich phase behavior.  The particles can agglomerate
at high densities, generally an irreversible process, but they may
also show a reversible liquid-vapor phase separation similar to the
one present in simple molecular liquids.  In order to prevent them
from simply falling out of solution, one needs some kind of repulsion
between the particles.  Introducing charged groups at the surface of
the colloid is one way to do that.  The large gain in entropy
following the dissociation of a vast number of counterions into
solution stabilizes the system, because an aggregation of colloids
into a small sub-volume would -- for reasons of global charge
neutrality -- also require the counterions to occupy this small volume
and thereby give up much entropy.  Of course, the final state of the
system is always a balance between energy and entropy, and if
electrostatic interactions are strong, they will ultimately overcome
entropy and lead to aggregation of the colloids
\cite{jensen98a,allahyarov98a,levin99a,messina00a,messina00b}.  The
resulting phenomenon of ``like charge attraction'' has received much
attention, but it is of course only mysterious if one forgets that the
entire system is neutral.  Admittedly, confusion persists about
whether such a phase separation could also happen within mean field
theory.  Even though rigorous proofs exist that PB theory will not
permit attraction between like charged macroions under reasonably
general circumstances \cite{neu99a,sader99a,trizac00a}, and that in a
cell model treatment the compressibility will be positive
\cite{tellez03a}, it has been claimed that an expansion of the free
energy of a charged colloidal suspension into zero-, one-, two-
etc. body terms will contain configuration independent volume terms,
which may drive a phase separation even though the pair terms are
purely repulsive \cite{roij97a,warren00a,dufreche03a}.  Since
unfortunately all these derivations rely on a linearization of PB
theory, which might render the findings as artifacts
\cite{grunberg01a,diehl01a,deserno02c,tamashiro03a}, the issue appears
to be open yet.

All these phenomena ultimately depend on the screening produced by the
ionic cloud, which in turn depends on the geometry of the system.  In
this regard, a charged spherical colloid differs from a charged rod in
two fundamental ways: the electrostatic potential and the spatial
extension.  The logarithmic potential present in the case of charged
cylinders leads to the phenomenon known as Manning condensation
\cite{oosawa71a,manning69a}.  If the line charge density exceeds a
critical threshold, a certain fraction will remain loosely associated
with the rod, even at infinite dilution, and renormalize the rod
charge.  A quantitative PB treatment of this provides a unique
criterion for defining the effective charge of the system, even at
finite densities \cite{lebret84a,deserno00a}.

The situation is different for charged spherical colloids, which lose
all their counterions in the limit of infinite dilution; thus, the
colloidal charge does not get renormalized.  Still, on often talks
about effective charges, which mimic the stronger condensation of
nonlinear theory within a linearized treatment
\cite{alexander84a,belloni98a,trizac02a,trizac03a,trizac03b,aubouy03a}.
That, however, is clearly not a physical but rather a formal
renormalization, necessitated by the simplified linear treatment, and
is thus a different story.

Another important difference between the spherical and the cylindrical
symmetry lies in the spatial extend.  If a charged rod is infinitely
long (as is usually assumed in theoretical treatments), the number of
counterions at any given distance from the rod is always infinite.  In
contrast, for a charged spherical colloid the number of counterions at
any distance is always finite, since of course there is no direction
along which the colloid is infinite.  Hence, fluctuations of the
radial charge density are more likely to be important in the spherical
case.

The systems we will consider here are strongly charged colloids with
point-like ions of some specific valence and no added salt inside a
spherical cell.  Since all the particles are limited to be within one
cell, correlations between \emph{different macroions} and between
microions belonging to \emph{different cells} are not present.  In our
treatment we will thus exclusively focus on questions regarding the
description of a \emph{single double layer}.  Furthermore, for
point-like ions the interpretation of our cutoff parameter, $a$, can
obviously no longer be the particle diameter.  We will introduce an
alternative prescription for $a$, based again on local density
considerations and keeping in mind that its entire purpose is to
prevent the functional from becoming unstable.

We also derive an approximated version of our correlation functional,
namely, its zero temperature limit.  It has the huge advantage that it
can be calculated analytically, while still predicting ion profiles
quite close to the full DHHC expression for a wide range of
parameters.  It also demonstrates the spirit of our stabilization
correction very directly.

Finally, we compare our predictions for ion profiles with Monte Carlo
simulations, in which we independently vary valence $v$ and plasma
parameter $\Gamma_{2d}=\sqrt{\pi\sigma\ell_\rB^2v^3}$, where $\sigma$
is the density of surface charges and $\ell_\rB$ is the Bjerrum
length.  It has been shown that beyond $\Gamma_{2d} \simeq 2.26$ the
force-distance curves between charged plates cease to be monotonic,
and beyond $\Gamma_{2d}\simeq 2.45$ attractions set in
\cite{moreira02a}.  These effects result from correlations between
different double layers (like, for instance, ion interlocking
\cite{jensen97a,deserno02b}), which we cannot account for, and it has
in fact been shown that they cannot be described within a local
density functional theory with a convex correlation correction
\cite{trizac00a}.  However, for the description of a
\emph{single} double layer the regime of applicability of our theory
is larger, even though it clearly must fail for too high coupling.

The paper is organized as follows. In Sec.\ II the DHHC correlation
functional is revisited and its zero temperature limit is introduced.
It is then applied as a local correlation correction to the problem of
screening of charged colloids in Sec.\ III. The case of point-like
ions is discussed in detail and the new expression for $a$ is
proposed.  Technical details of the simulations are described in Sec.\
IV.  The results of the simulations, full theory and zero temperature
limit are compared in Sec.\ V, and we end with our conclusions Sec.\
VI.

%%%%%%%%%%%%%%%%%%%%%%%%%%%%%%%%%%%%%%%%%%%%%%%%%%%%%%%%%%%%%%%%%%%%%%%%%%%%%%

\section{The Debye-H\"uckel Hole-Cavity (DHHC) Theory Revisited}

The one component plasma consists of $N$ identical point-particles of
valence $v$ and (positive) unit charge $q$ inside a volume $V$ with a
uniform neutralizing background of charge density $-v q n_\rB$ and
dielectric constant $\varepsilon$.  As a first approximation the free
energy of this system can be derived in the framework of the
Debye-H\"uckel approach.  Then, the electrostatic potential $\psi$
created by some ion, fixed at the origin for instance, and all its
surrounding ions satisfies the spherically symmetric Poisson equation
$\nabla^2 \psi(r) = \psi''(r) + \frac{2}{r}\,\psi'(r) = -4 \pi
\rho(r)/\varepsilon$.  The charge density has a contribution from the
central ion, $vq\delta(\VECr)$, a contribution from the surrounding
ions which are distributed -- within mean field theory! -- according
to the Boltzmann factor $n_\rPB(\VECr)=vqn_\rB \exp\{-\beta v q \psi
(r)\}$, and finally from the charged background.  Inserting this into
the Poisson equation and linearizing the exponential yields the
linearized Poisson-Boltzmann equation
\begin{equation}
  \label{eq:LPB}
  \psi''(r) + \frac{2}{r}\,\psi'(r) =
  \kappa^2 \psi - \frac{4\pi}{\varepsilon}vq\delta(\VECr) \ ,
\end{equation}    
where $\kappa\equiv\sqrt{4\pi\ell n_\rB}$ is an inverse screening
length, $\ell=\ell_\rB v^2$, $\ell_\rB=\beta q^2/\varepsilon$ is the
Bjerrum length, and $\beta=1/k_\rB T$ is the inverse thermal energy.

The solution of Eqn.~(\ref{eq:LPB}) is the well known expression
$\psi(r) = vq\,\re^{-\kappa r}/\varepsilon r$.  However, the problem with
Debye-H\"uckel theory is that the condition for linearization is
obviously not satisfied for small $r$, where the potential is large.
Indeed, since all ions have the same sign of charge, this implies that
the particle density becomes \emph{negative} and finally diverges at
the origin.  This defect was overcome by the Debye-H\"uckel-Hole
theory \cite{nordholm84a}, which artificially postulates a correlation
hole of radius $h$ around the central ion into which no other ions are
allowed to penetrate.  In this case the charge density is given by
%==============================
\begin{equation}
\label{eq:rho_DHH}
  \rho(r) \; = \;
    \left\{\begin{array}{r@{\quad:\quad}l}
      vq\,(\delta(\VECr) - n_\rB)  & r \le h \\[2ex]
      -\displaystyle\frac{\varepsilon \kappa^2}{4\pi} \psi(r) & r > h
    \end{array}
  \right.
  \ .
\end{equation}
%===============================

The hole size $h$ is fixed by excluding particles from a region where
their Coulomb energy is larger than $k_\rB T$, which gives $1+\kappa
h=(1+3\kappa\ell)^{1/3}$.  Once the potential at the position of the
central ion is known, the electrostatic contribution to the free
energy density, $f_\rDHH(n)$, can be obtained by the Debye charging
process \cite{debye23a,penfold90a}.

The simple Debye-H\"uckel-Hole analysis of the one-component plasma
theory offers considerable insight into ionic systems and is in good
agreement with Monte-Carlo simulations \cite{brush66a} when
fluctuations on the charge density are not relevant \cite{tamashiro99a}.
In principle, one can attempt to include such fluctuations by applying
the bulk density-functional theory in a local way.  The basic idea is
to obtain the density distribution via functional minimization of the
free energy
%==============================
\begin{equation}
  \label{eq:F_OCP}
  F_\rOCP[n(\VECr)] = F_\rPB[n(\VECr)] + \int\!\infd^3r \; f_\corr\big(n(\VECr)\big) \ .
\end{equation}
%===============================
The first part, the PB free energy
%==============================
\begin{equation}
  \label{eq:F_PB}
  F_\rPB[n(\VECr)]
  =
  \int \! \infd^3r \;
   \Big \{ k_\rB T n(\VECr) \big[ \ln \big( n(\VECr) V_\rp \big) -1 \big]\, + \, f_\rel \Big\} \ ,
\end{equation}
%==============================
contains the entropy of the mobile ions, the interaction of the small
ions with the macroion potential and the mean-field interaction
between the counterions.  Here $V_p$ is the particle volume.  The
expression $f_\corr$ in Eqn.~(\ref{eq:F_OCP}) accounts for the
correlation between the mobile ions.  The ion distribution can be
derived by minimization of Eqn.~(\ref{eq:F_OCP}) under the constraint
of charge neutrality.  Unfortunately, this variational process does
not lead to a well defined density profile if one uses $f_\rDHH(n)$ as
the correlation correction $f_\corr$.  The reason is that $f_\rDHH(n)$
is a concave function beyond $n^\star\approx 7.86/\ell^3$ and
asymptotically behaves as $-n^{4/3}$.  Since there is no stabilizing
homogeneously charged background but rather a concentration of
opposite charge on the macroion surface, this favors the development
of a distribution in which all the ions sit on the surface of the
macroion.
 
The instabilities present in the DHH approach can be properly overcome
by recognizing that the failure of this model is due to the too strong
requirement of local charge neutrality imposed by the local density
approximation: A local fluctuation leading to an increase of particle
density implies a corresponding increase in background density.
Therefore, the fluctuation is not suppressed by an increase in
repulsive Coulomb interactions but quite on the contrary favored by
its decrease.  To circumvent the instabilities occurring at high
densities, we proposed recently a simple solution in which one
excludes the neutralizing background from a cavity of radius $a$
placed around the central ion (for details of the derivation of the
model see Ref. \cite{barbosa00a}).  In this case, the charge density
can be split in three different regions, namely
%============================== 
\begin{equation}
  \label{eq:rho_DHHC}
  \rho(r) \; = \; \left\{
    \begin{array}{r@{\quad:\quad}l}
      vq\delta(\VECr) & 0 \le r < a \\
      -vqn_\rB & a \le r < h \\
      -\displaystyle\frac{\varepsilon \kappa^2}{4\pi} \psi(r) & h \le r\;
    \end{array}
  \right.
  \ ,
\end{equation}
%===============================
where the hole size $h$ is chosen such as to yield the same screening
(i.e., the same amount of charge within $h$) as the DHH theory, which
results in
%==============================
\begin{equation}
  \label{eq:h_DHHC}
  \kappa h = \big[(\omega-1)^3+(\kappa a)^3\big]^{1/3} \ ,
\end{equation}
%==============================
with $\omega=(1+3\kappa\ell)^{1/3}$.  Using this prescription for $h$,
the free energy is obtained by Debye-charging the fluid:
%==============================
\begin{eqnarray}
  \label{eq:F_DHHC}
  \frac{\beta f_\rDHHC}{n_\rB}
  & = &
  \frac{(\kappa a)^2}{4} - 
  \int_{1}^{\omega} \infd\overline{\omega} \, \Big\{
  \frac{\overline{\omega}^2}{2(\overline{\omega}^3-1)}\Omega(\overline{\omega})^{2/3}
  \nonumber \\
  & & \quad
  +\frac{\overline{\omega}^3}{(1+\Omega(\overline{\omega})^{1/3})
    (\overline{\omega}^2+\overline{\omega}+1)} \Big\} \ ,
\end{eqnarray}
%==================================
where
%==================================
\begin{equation}
  \label{eq:Omega}
  \Omega(\overline{\omega}) = (\overline{\omega}-1)^3
  +\frac{(\kappa a)^3}{3 \kappa\ell}\,(\overline{\omega}^3-1) \ .
\end{equation}
%==============================

Since the DHHC free energy is a convex function of density, $f_\rDHHC$
can thus be used to account for correlations within a local density
approximation.

%%%%%%%%%%%%%%%%%%%%%%%%%%%%%%%%%%%%%%%%%%%%%%%%%%%%%%%%%%%%%%%%%%%%%%%%%%%%%%

\subsection{The Zero Temperature Limit \rDHHCz}

The fact that the integral in Eqn.~(\ref{eq:F_DHHC}) has to be solved
numerically obstructs a direct view on how thermodynamic stability is
actually restored.  Luckily, the crucial point can already be seen by
focusing on the limit of zero temperature.  In this case
Eqn.~(\ref{eq:h_DHHC}) gives the expression
\begin{equation}
  h = (3/4\pi n_\rB + a^3)^{1/3}
  \label{eq:h_DHHC0}
\end{equation}
for the correlation hole of the DHHC theory.  This
conveniently implies the potential to vanish outside $h$.  In other
words, the region $a<r<h$ contains the right amount of background
charge to exactly neutralize the central ion, and it is appropriate to
refer to this limit as ``complete screening''.  The potential in the
two other regions then simplifies considerably:
%==============================
\begin{widetext}
\begin{eqnarray}
  \hspace*{-2ex}
  \psi(r) & = &  \frac{v \, q}{4\pi\epsilon r}\times
  \left\{
    \begin{array}{l@{\quad:\quad}l}
      1 \; + \; \D\frac{3r}{2a}\Big(\nhat_\rB -
      \frac{a}{h}(1+\nhat_\rB)\Big)  & 0 \le r < a \\[2ex]
      1 \; + \; \nhat_\rB\Big(1+\D\frac{r^3}{2a^3}\Big) \; - \; 
      \D\frac{3r}{2h}\Big(1+\nhat_\rB\Big)
      & a \le r < h
    \end{array} 
  \right.
  \ ,
\end{eqnarray}
\end{widetext}
%==============================
with the dimensionless scaled density $\nhat_\rB$ given by $\nhat_\rB
= \frac{4}{3}\pi a^3 \, n_\rB$.  After the Debye charging process one
obtains the following closed expression for the excess free energy
density:
%==============================
\begin{equation}
  \frac{\beta f^{(0)}_\rDHHC}{n_\rB}
  =
  \frac{3\ell}{4 a}
  \left\{\nhat_\rB - (1+\nhat_\rB)^{2/3}\,\nhat_\rB^{1/3}\right\} \ .
  \label{eq:f_DHHC0}
\end{equation} 
%==============================

Note that the limits $a\rightarrow 0$ and $n_\rB \rightarrow \infty$
do \emph{not} commute: For high densities, $\beta f^{(0)}_\rDHHC$
scales asymptotically like $-\ell n_\rB / 2a$, i.e., linear with
density.  However, in the limit $a\rightarrow 0$
Eqn.~(\ref{eq:f_DHHC0}) becomes
%==============================
\begin{equation}
  \lim_{a\rightarrow 0} \beta f^{(0)}_\rDHHC
  = 
  -\ell \left(\frac{9\pi}{16}\right)^{1/3}n_\rB^{4/3} \ ,
  \label{eq:f_DHHC0_a0}
\end{equation}
%==============================
and this concave scaling with density prevents it from being used
within a local density approximation.  The zero temperature limit thus
demonstrates in a clear way the key role played by the cavity of size
$a$, which excludes the uniform background from the vicinity of the
central ion.

%%%%%%%%%%%%%%%%%%%%%%%%%%%%%%%%%%%%%%%%%%%%%%%%%%%%%%%%%%%%%%%%%%%%%%%%%%%%%%

\subsection{How to choose a proper value for $a$}

Before applying this strategy to various valences and ionic strengths,
we need to specify the parameter $a$.  If the counterions have a
diameter $d$, no other charge should be found at a distance $r<d$.
Therefore, in the spirit of the Debye-H\"uckel theory, we tentatively
interpreted $a$ in Ref. \cite{barbosa00a} as the ion diameter.  This
choice has led to an excellent agreement with simulations when applied
to rod-like polyelectrolytes \cite{barbosa00a} with mono-, di-, and
trivalent counterions (and no added salt), but it is of course
infeasible for point ions.  In the following we suggest an alternative
way to choose a value for $a$ which is independent of excluded volume
arguments, and show that this choice yields a good description of our
Monte Carlo results and trends.

We already mentioned the crucial role played by $a$ in maintaining the
free energy convex.  We also have seen in our discussion of the
zero-temperature case that this is achieved because $a$ in
Eqn.~(\ref{eq:h_DHHC0}) balances the length $(3/4\pi n_\rB)^{1/3}$,
which is basically the mean distance between ions.  One could thus try
to selfconsistently choose $a$ proportional to the local ion distance,
but this would be unsuccessful: The balance would not work, since each
density increase would shrink $a$ proportionally, and the collapse
could not be stopped.  One thus needs a length which is local and
\emph{somehow} related to the ion density -- but which does \emph{not}
change as the local ion density changes.  This suggests to pick the
average distance between ions \emph{as predicted by PB theory}:
$a=(3/4\pi n_\rPB(r))^{1/3}$.  Our density functional then quite
naturally emerges as a next order correction to the mean field result.

After these general considerations on the cutoff $a$, let us continue
with a practical remark.  Far away from the charged surface the ion
density is always quite low, correlations are weakly developed, and
the precise value of $a$ is immaterial.  In fact, we only ever need a
stabilizing cutoff close to the charged surface, where the ion density
is largest.  This suggests the following simplification: Instead of
using a cutoff function $a(n_\rPB(r))$ depending on the local PB
density, we pick a constant $a$ from a worst-case scenario, namely,
the value which it has at contact.  This then finally yields the
following prescription for $a$:
%==============================
\begin{equation}
 a = \Big(\frac{3}{4 \pi  n_\rPB(r_0)}\Big)^{1/3} \ .
 \label{eq:a2}
\end{equation}
%==============================
In fact, since the cutoff will become important in the regime of
strong correlations, we could even replace the contact density
$n_\rPB(r_0)$ by its limiting value $2\pi\ell_\rB\sigma^2$, where
$\sigma$ is the density of surface charges \cite{wennerstroem82a}.  We
then find
\begin{equation}
  \frac{a}{\ell}
  \stackrel{\text{strong coupling}}{\longrightarrow}
  \left(\frac{3}{8\pi^2\sigma^2\ell_\rB^4v^6}\right)^{1/3}
  \; = \;
  0.721 \, \Gamma_{2d}^{-4/3} \ ,
  \label{eq:a_lim}
\end{equation}
where
%==============================
\begin{equation}
  \Gamma_{2d}=\sqrt{\pi\sigma\ell_\rB^2v^3}
  \label{eq:Gamma}
\end{equation}
%==============================
is the $2d$ plasma coupling parameter \cite{baus80a}.  Formula
(\ref{eq:a_lim}) nicely demonstrates that in this limit the cavity
size, measured in the appropriate length scale $\ell$ (see also the
scaling discussion in the Appendix), is simply another measure of the
coupling strength.

%%%%%%%%%%%%%%%%%%%%%%%%%%%%%%%%%%%%%%%%%%%%%%%%%%%%%%%%%%%%%%%%%%%%%%%%%%%%%%
%%%%%%%%%%%%%%%%%%%%%%%%%%%%%%%%%%%%%%%%%%%%%%%%%%%%%%%%%%%%%%%%%%%%%%%%%%%%%%

\section{Application to the spherical cell model}\label{sec:application}

Charged spherical colloids are common and well characterizable systems
for studying many electrostatic phenomena in pure culture, and they
can often serve as simplified models for more complicated systems like
polyelectrolytes or proteins.  Solutions containing such charged
structures are indeed complicated to describe due to the long-range
nature of the Coulombic interactions.  However, as long as these long
range forces are repulsive, the colloids will create large correlation
holes (``cells'') around themselves which are void of other colloids.
In a first approximation one can then decouple the macroion
interactions and concentrate on what's going on within a single
correlation hole---an approach which is termed ``cell model''
\cite{katchalsky71a}.  The cell picture is known to give a good
approximation for many realistic systems, and most of the physics of
the system is determined by the screening of the macroion by the
microions inside a cell.  As a test case for our theory we shall
therefore consider a charged spherical colloid of radius $r_0$
containing $Z$ charged groups, which are neutralized by point-like
ions of valence $v$.  This macroion is embedded in the center of a
spherical cell of radius $R$, corresponding thus to a volume fraction
$\phi=(r_0/R)^3$ of colloids.

The thermodynamic behavior of the colloidal system is determined by
the distribution of mobile ions around the macroion. This distribution
is obtained by minimization of the free energy functional,
Eqn.~(\ref{eq:F_OCP}).  For the colloidal system, the interaction of
the small ions with the macroion and the mean-field interaction
between the counterions are given by
%==============================
\begin{equation}
  f_\rel = \frac{1}{2} v q n(\VECr)\Big(\psi(\VECr)+\psi_{\text{fix}}(\VECr)\Big) \ ,
  \label{eq:f_el}
\end{equation}
%==============================
where $\psi(\VECr)$ is the total electrostatic potential at position
$\VECr$ and $\psi_{\text{fix}}(\VECr)=-Zq/\varepsilon r$ is the
potential due to the charged macroion alone.  The inter-particle
correlations are taken into account by employing $f_\corr = f_\rDHHC$.
The minimization itself is accomplished by numerically solving the
corresponding Euler-Lagrange-equation.  Special care had to be taken
to obtain a sufficient accuracy of the rapidly varying density
profiles close to the colloid surface.

In the following we will concentrate on two observables.  The first is
the integrated fraction of ions within a radial distance $r$ from the
colloid center, which is given by
%==============================
\begin{equation}
  P(r) = \frac{1}{Z}\int_{r_0}^{r} \! \infd\bar{r} \; 4\pi \bar{r}^2 \, v q  n(\bar{r}) \ .
  \label{eq:P}
\end{equation}
%==============================
We will sometimes plot $P$ as a function of $-1/r$, which is the Green
function of the spherical Laplacian.  This will visually expand the
region close to the colloid, but it also has practical advantages when
estimating the amount of closely associated ions, see e.g.\
Ref.~\cite{belloni84a,belloni98a}.

Measuring all lengths in the full partition function of the cell model
in units of $\ell=\ell_\rB v^2$ reveals that the distribution function
$P(r)$ is invariant under a rescaling which keeps the number of
counterions $N=Z/v$, the reduced colloid size $r_0/\ell$, and the
volume fraction $\phi=(r_0/R)^3$ constant (see Appendix).  The same
holds for PB theory, and it is also true for DHHC theory.  In the
latter case this not only relies on the form of the DHHC free energy
correction (\ref{eq:F_DHHC}), but also on our particular choice of
$a$.  This invariance property is thus a further support for
Eqn.~(\ref{eq:a2}).

The second observable we look at is the osmotic pressure $\Pi$.  For
PB like free energy functionals with an additional density term --
like our $f_\corr$ -- it is given by \cite{tellez03a}
\begin{equation}
  \beta \Pi =
  \Big[n + n\frac{\partial f_\corr(n)}{\partial n} - f_\corr(n)\Big]_{n=n(R)} \ .
  \label{eq:pressure}
\end{equation}
For the PB case, $f_\corr \equiv 0$, this reduces to the well known
fact that the pressure is given by the boundary density
\cite{marcus53a}.  Since this result actually holds \emph{rigorously}
for the full restricted primitive model \cite{wennerstroem82a}, one
could also argue that DHHC theory is an approximate way to calculate
the boundary density, and then calculate the pressure from
$\beta\Pi=n(R)$, i.e., leave out the additional term $nf'-f$.  This
would lead to a different result, reminding us that selfconsistency
and consistency with other rigorous results cannot generally be
achieved.  We will always use the internally consistent equation
(\ref{eq:pressure}) for our pressure calculations.

Inserting the DHHC expression (\ref{eq:F_DHHC}) for $f_\corr$, we find
\begin{eqnarray}
  \label{eq:p_DHHC}
  \frac{\beta \Pi_\rDHHC}{n(R)}
  & = &
    1 + \frac{(\kappa a)^2}{4} -\frac{1}{6}
    \int_{1}^{\omega} \! \infd\overline{\omega} \; \Big[
    \frac{\Phi(\overline{\omega})}{\Omega(\overline{\omega})^{1/3}} 
  \nonumber \\
  & & \hspace*{-4em} +\frac{2\overline{\omega}-1}{(1+\Omega(\overline{\omega})^{1/3})} 
    -\frac{(\overline{\omega}^2-\overline{\omega})\Phi(\overline{\omega})}
    {(\Omega(\overline{\omega})^{1/3}+\Omega(\overline{\omega})^{2/3})^2 } \Big] \ ,
\end{eqnarray}
where $\Phi(\overline{\omega}) = (\overline{\omega}-1)^2 + (\kappa
a)^3\overline{\omega}^2/\kappa\ell$.  In the zero temperature limit
\rDHHCz\ this simplifies considerably.  A closed expression can easily
be derived by combining Eqns.~(\ref{eq:f_DHHC0}) and
(\ref{eq:pressure}):
\begin{eqnarray}
  \frac{\beta \Pi_\rDHHC^0}{n(R)}
  & = &
  1 + \frac{\ell}{4a}\nhat
      \Big\{3-2\,\big(1+\nhat^{-1}\big)^{-1/3}
  \nonumber \\
  & & \qquad -\big(1+\nhat^{-1}\big)^{2/3}\Big\} \label{eq:p_DHHC_0} \\
  & = &
  1 - \frac{\ell}{4a}\nhat^{1/3} \Big\{1 - 3\,\nhat^{2/3}
   + {\mathcal O}\big(\nhat\big) \Big\} \ ,
  \label{eq:CSpressure}
\end{eqnarray}
where $\nhat\equiv\frac{4}{3}\pi a^3 n(R)$.  Observe that the
contribution originating from the $nf'-f$ term is negative for all
densities.

%%%%%%%%%%%%%%%%%%%%%%%%%%%%%%%%%%%%%%%%%%%%%%%%%%%%%%%%%%%%%%%%%%%%%%%%%%%%%%
%%%%%%%%%%%%%%%%%%%%%%%%%%%%%%%%%%%%%%%%%%%%%%%%%%%%%%%%%%%%%%%%%%%%%%%%%%%%%%

\section{Simulational details}

The systems we study consist of a spherical macroion of radius $r_0$
and (negative) central charge $-Zq$.  Electroneutrality is ensured by
the presence of $N=Z/v$ point-like counterions of valence $v$,
confined inside an impermeable spherical cell of radius $R$.  This
also fixes the colloid volume fraction to $\phi=(r_0/R)^3$.  No
additional salt is added.  The dielectric constant $\varepsilon$ is
assumed to be uniform throughout the system, such that no image forces
\cite{messina002f} occur.  Our choices for the system parameters can
be found in Table~\ref{systems}

%%%%%%%%%%%%%%%%%%%%% Table 1
\begin{table}[tbp]
\begin{ruledtabular}
\begin{tabular}{cccccddd}
System & $Z$ & $N$ & $v$ & $r_0/\ell$ & \multicolumn{1}{c}{$\Gamma_{2d}$} & \multicolumn{1}{c}{$n_\rPB(r_0)\,\ell^3$} & \multicolumn{1}{c}{$a/\ell$} \\[.5ex] \hline
  1    & 100 & 100 & 1 & 2          & 2.5                               & 20.77                                         & 0.23  \\
  2    & 120 & 120 & 1 & 5.477      & 1                                 & 0.4090                                        & 0.836 \\
  3    & 120 & 120 & 1 & 2.739      & 2                                 & 8.275                                         & 0.307 \\
  4    & 120 & 120 & 1 & 1.826      & 3                                 & 45.07                                         & 0.174 \\
  5    & 120 &  60 & 2 & 1.937      & 2                                 & 7.526                                         & 0.317 \\
  6    & 120 &  40 & 3 & 1.581      & 2                                 & 6.976                                         & 0.324     
\end{tabular}
\caption{The parameters of the simulated systems.  The volume fraction
was always chosen as $\phi=(r_0/R)^3=0.8\%$.}\label{systems}
\end{ruledtabular}
\end{table}
%%%%%%%%%%%%%%%%%%%%%

Standard canonical MC simulations following the Metropolis scheme
\cite{metropolis53a} were employed to sample the ion distributions.  After an
initial equilibration time of 200\,000 MC steps, where we attempted to
move every ion once to a new position, we sampled the system for
$1.3-2 \times 10^{6}$ MC steps, producing 1300--2000 configurations
for analysis.  We will measure energies in units of $k_\rB T$ and use
the coupling length $\ell=\ell_\rB v^2$ as our unit of length (for
monovalent ions under aqueous conditions and room temperature we would
have $\ell = 7.14\,\text{\AA}$, and the unit of concentration becomes
$\ell^{-3}=4.56\,\text{M}$).  In the following we will present MC
results for the integrated ion distribution, Eqn.~(\ref{eq:P}), and
for the pressure, Eqns.~(\ref{eq:p_DHHC}) and (\ref{eq:p_DHHC_0}).

%%%%%%%%%%%%%%%%%%%%%%%%%%%%%%%%%%%%%%%%%%%%%%%%%%%%%%%%

\section{Comparison between Simulations and DHHC theory}

%%%%%%%%%%%%%%%%%%%%%%%%%%%%%%%%%%%%%%%%%%%%%%%%%%%%%%%%

\subsection{Ion distribution functions}

\begin{figure}
\includegraphics[scale=0.84]{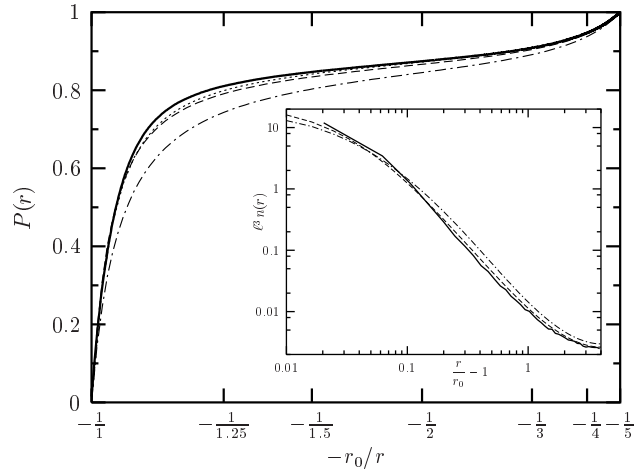}
\caption{Counterion distribution function $P(r)$ for system 1 (see
  Tab.~\ref{systems}).  The solid curve is the result of the MC
  simulation, while the dash-dotted curve is the prediction from PB
  theory.  The inset shows the local density $n(r)$.  The increase in
  the counterion condensation due to correlations is well captured by
  the \rDHHC\ theory (dashed curve) and its zero temperature limit
  \rDHHCz\ (dotted curve).  The difference in $n(r)$ between the
  latter two is invisible on the chosen scale, and only \rDHHC\ is
  shown.}
\label{fig:Z=100}
\end{figure}

Figure~\ref{fig:Z=100} shows the integrated charge distribution $P(r)$
for system 1 from Tab.~\ref{systems}.  The solid curve is the result
from the MC simulation, and it lies distinctly above the PB result
(dash-dotted curve), indicating a stronger condensation of ions due to
correlations neglected in PB theory.  Most of this enhancement of ion
localization close to the colloid is captured by \rDHHC\ theory
(dashed curve) or its zero temperature limit \rDHHCz.  This is also
evident from the local density $n(r)$, which relatively to PB is
enhanced at close proximity to the colloid, while it drops below PB at
the outer cell boundary.  From what we have said in section
\ref{sec:application} this also indicates that the pressure will be
lower, and this is indeed what we shall find (see below).

It should be noted that the 2d plasma parameter $\Gamma_{2d}=2.5$ is
already slightly beyond the point where attractions between two planes
would arise \cite{moreira02a}.  We should not expect \rDHHC\ theory to
work for significantly higher plasma parameters, since it cannot
account for effects like attractions \cite{trizac00a}.  However, we
want to point out that here we only aim at properties of a
\emph{single} electrostatic double layer and not at phenomena arising
from the interaction between two of them, and in fact the agreement
seen in Fig.~\ref{fig:Z=100} is very encouraging.  It is also quite
pleasing that the significantly simpler zero temperature limit
\rDHHCz\ from Eqn.~(\ref{eq:f_DHHC0}) yields essentially the same result
as the full \rDHHC\ theory.

Due to the scaling invariance of the partition function discussed in
the Appendix, a system with e.g. divalent ions and $Z=200$ or
trivalent ions and $Z=300$ (and properly rescaled Bjerrum lengths
$\ell_\rB \rightarrow \ell_\rB/v^2$) shows exactly the same
distribution function (not shown).

\begin{figure}
\includegraphics[scale=0.82]{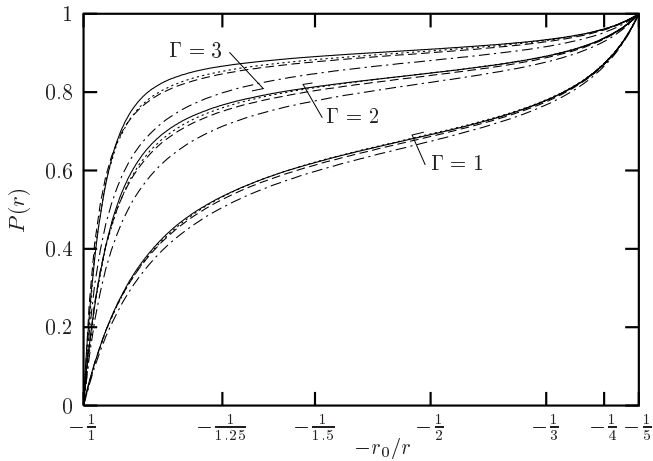}
\caption{Counterion distribution function $P(r)$ for systems 2, 3,
  and 4 from Tab.~\ref{systems}.  The line styles are the same as in
  Figure \ref{fig:Z=100}, the counterions are monovalent, and the
  value of the plasma-parameter $\Gamma_{2d}$ is indicated.
  }\label{fig:Gamma_scan}
\end{figure}

In Figure \ref{fig:Gamma_scan} we show distribution functions $P(r)$
for systems 2--4 from Tab.~\ref{systems}.  These have monovalent
counterions and only differ in their value of the plasma parameter
$\Gamma_{2d}$.  Clearly, a larger plasma-parameter leads to an
increased condensation (the curves are shifted up)---an effect which
naturally is already present in PB theory.  However, apart from this,
at a larger plasma parameter the influence of correlations becomes
more important, and therefore the \emph{deviation} between the PB
prediction and the MC result increases for increasing $\Gamma_{2d}$,
which is also clearly seen in Fig.~\ref{fig:Gamma_scan}.  Again, this
effect is well captured by \rDHHC\ theory, which is always much closer
to the MC data than to the PB result, even though its accuracy
diminishes as $\Gamma_{2d}$ becomes large.

\begin{figure}
\includegraphics[scale=0.82]{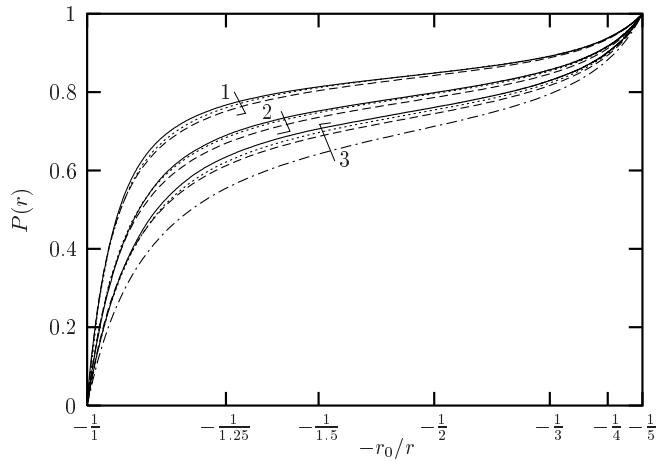}
\caption{Counterion distribution function $P(r)$ for systems 3, 5,
  and 6 from Tab.~\ref{systems}.  The line styles are the same as in
  Figure \ref{fig:Z=100}, the plasma parameter is $\Gamma_{2d}=2$, and
  the value of the counterion valence is indicated.  For clarity, the
  PB curve is only shown for $v=3$.}
\label{fig:v_scan}
\end{figure}

In Fig.~\ref{fig:v_scan} we show a ``complementary'' scan, in which we
fixed the value of the plasma-parameter $\Gamma_{2d}=2$, but changed
the counterion valence (systems 3, 5, and 6 from Tab.~\ref{systems}).
Maybe surprisingly, an \emph{increase} in valence leads to a
\emph{decrease} in condensation if it happens at constant
plasma-parameter and colloid charge.  If we had changed $v$ from 1 to
2 and simultaneously replaced $\ell_\rB\rightarrow \ell_\rB/4$
\emph{and} $Z\rightarrow 2Z$, the plasma parameter would also have
remained unchanged, but due to the scaling property of the partition
function that would actually have been true for the whole distribution
function.  Instead, we have reduced $\ell_\rB \rightarrow
\ell_\rB/2^{3/2} \approx \ell_\rB/2.83$ (i.e., a little less strongly),
but have failed to increase $Z$.  The net result is that condensation
drops slightly.  However, since the plasma parameter, which is the
best indicator for the strength of correlations, remains the same, the
\emph{deviation} between PB theory and MC simulation are always about
the same (not shown in the Figure).  And as a consequence, the
deviation between \rDHHC\ theory, which approximately accounts for
correlations, and the MC simulation, which captures them all, is about
the same in all three cases, and actually not very big.

%%%%%%%%%%%%%%%%%%%%%%%%%%%%%%%%%%%%%%%%%%%%%%%%%%%%%%%%

\subsection{Osmotic Pressure}

Another strategy to check how successful our approach captures
correlations is to compute the osmotic pressure.  In real systems this
pressure will depend on correlations between ions of \emph{different}
cells, something which neither our theory nor actually our simulation
(of a single colloid!) takes into account.  So in the following by
``pressure'' we do not, strictly speaking, refer to the bulk pressure
of a colloidal suspension at some given volume fraction, but only to
the pressure exerted on the rigid wall at $r=R$ of our cell model.

\begin{figure}
\includegraphics[scale=0.82]{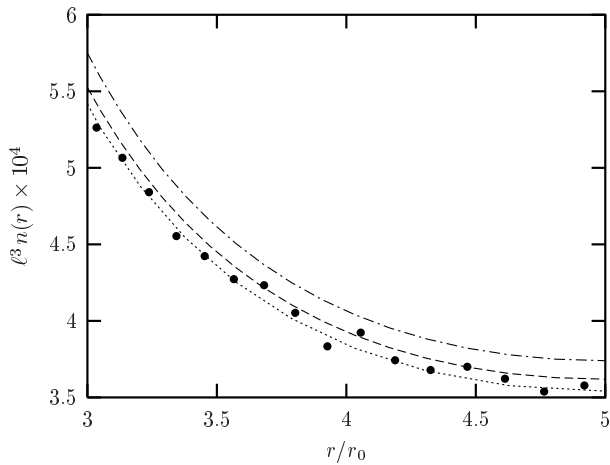}
\caption{Counterion density  close to the cell boundary for
  system 2.  The dots denote the results of the MC simulation, the
  other line styles are the same as in Fig.~\ref{fig:Z=100}.}\label{fig.c1}
\end{figure}

Within the simulations, the pressure is given by the contact density
at $r=R$, which was obtained by fitting the MC density profile close
to the cell boundary to a quadratic expression $n(r)=c_1+c_2(r-R)^2$.
An example for how the simulated densities compare to the PB
approximation, our analytic DHHC approach, and its simpler zero
temperature limit \rDHHCz, can be found in Fig.~\ref{fig.c1}.

%%%%%%%%%%%%%%%%%%%%% Table 2
\begin{table*}
\begin{ruledtabular}
\begin{tabular}{cccccccc}
Sys. & $\beta \Pi_\rPB\ell^3$ & $\beta \Pi_\rDHHC\ell^3$ & $\beta \Pi_\rDHHC^0\ell^3$ & $\beta \Pi_\rMC\ell^3$   & $\Pi_\rPB/\Pi_\rMC$ & $\Pi_\rDHHC/\Pi_\rMC$ & $\Pi_\rDHHC^0/\Pi_\rMC$ \\[.5ex] \hline
  1  & $2.98 \times 10^{-3}$ & $2.56 \times 10^{-3}$ & $2.39 \times 10^{-3}$ & $2.53(3) \times 10^{-3}$ & 1.18 & 1.01 & 0.94 \\
  2  & $3.74 \times 10^{-4}$ & $3.58 \times 10^{-4}$ & $3.44 \times 10^{-4}$ & $3.55(4) \times 10^{-4}$ & 1.05 & 1.01 & 0.97 \\
  3  & $1.58 \times 10^{-3}$ & $1.42 \times 10^{-3}$ & $1.33 \times 10^{-3}$ & $1.38(3) \times 10^{-3}$ & 1.14 & 1.03 & 0.96 \\
  4  & $3.61 \times 10^{-3}$ & $3.02 \times 10^{-3}$ & $2.81 \times 10^{-3}$ & $2.95(5) \times 10^{-3}$ & 1.22 & 1.02 & 0.95 \\
  5  & $3.09 \times 10^{-3}$ & $2.72 \times 10^{-3}$ & $2.53 \times 10^{-3}$ & $2.63(6) \times 10^{-3}$ & 1.17 & 1.03 & 0.96 \\
  6  & $4.10 \times 10^{-3}$ & $3.96 \times 10^{-3}$ & $3.71 \times 10^{-3}$ & $3.83(9) \times 10^{-3}$ & 1.07 & 1.03 & 0.97 \\
\end{tabular}
\caption{The values of the various pressures (in units $k_\rB T/\ell^3$) for
the systems 1--6.  The MC errors have been conservatively estimated
from the fluctuations of the measured density around the fit close to
the cell boundary.  The last three columns display the ratio between
the theoretical and MC values, illustrating which theories over- or
underestimate the pressure, and by how much.}\label{pressure}
\end{ruledtabular}
\end{table*}
%%%%%%%%%%%%%%%%%%%%%

Table \ref{pressure} shows the predictions for the pressure in the
case of point-like ions given by PB-, \rDHHC-, and \rDHHCz-theory, as
well as by MC simulations.  In the case of PB-theory and MC the
pressure is simply the density at the outer boundary, while for the
DHHC approach we employ Eqn.~(\ref{eq:p_DHHC}) and for \rDHHCz\
Eqn.~(\ref{eq:CSpressure}).  These data, as well as Fig.~\ref{fig.c1},
demonstrate that -- as anticipated -- the simulated pressures lie
below the PB prediction.  This decrease in pressure is rather
accurately captured by our functional $f_{\rDHHC}$ and by its zero
temperature limit, $f^{(0)}_{\rDHHC}$, with the MC result lying
significantly below PB and (in these cases) below \rDHHC\ and above
\rDHHCz.  The difference between the two correlation corrected
approaches is consistent with the idea that entropic effects neglected
in $f^{(0)}_{\rDHHC}$ would push ions away from the macroions, or in
other words, that the zero temperature limit implies stronger
correlations than the regular \rDHHC\ theory and therefore yields even
lower pressures.

%%%%%%%%%%%%%%%%%%%%%%%%%%%%%%%%%%%%%%%%%%%%%%%%%%%%%%%%%%%%%%%%%%%%%%%%%%%%%%
%%%%%%%%%%%%%%%%%%%%%%%%%%%%%%%%%%%%%%%%%%%%%%%%%%%%%%%%%%%%%%%%%%%%%%%%%%%%%%

\section{Conclusions}

In this paper we showed how to apply our previously proposed local
density functional approach based on a stable correlation correction
to a spherical macroion confined in a spherical cell.  One of the
crucial parameters in this theory is the size $a$ of the exclusion
cavity of the background charge density.  For point-like ions, we
suggest to associate the exclusion region with the mean distance
between ions as predicted by PB theory, and for simplicity use the
value present at colloidal contact.

By going to the zero temperature limit we were able to derive an even
simpler free energy functional $F^{(0)}_\rDHHC$, which is almost as
good as the full \rDHHC\ theory, but much easier to handle.  We also
derived exact expressions for the osmotic pressure in this system. We
successfully compared our predictions to simulations of the same model
and compared the integrated counterion density and the osmotic
pressure values for two complementary ``scans'' of the coupling
strength, namely valence and plasma parameter.  We demonstrated that
our local density functional approach based on a stable correlation
correction leads to a major improvement over the PB prediction.

\section*{Acknowledgments}

This work has been supported by the Brazilian agency, CNPq (Conselho
Nacional de Desenvolvimento Cient\'{\i}fico e Tecnol\'ogico) and the
German Science Foundation (DFG) through SFB 625, TR6, Ho-1108/11-1 and
De$\,$775/1-2.  MCB would like to thank Prof.\ K.\ Kremer for the
hospitality during her stay in Mainz.

\appendix

\section*{Appendix}

The canonical partition function $\CALZ$ of the colloid
surrounded by its counterion is given by:
%==============================
\begin{equation}
  {\cal Z} = 
  \int_{}^{} \prod_{i=1}^{N} \frac{\infd^{3}p_i \, \infd^{3}r_i}{h^{3N}N!} \,\re^{-\beta {\cal H}} \ ,
 \label{eq:ap1}
\end{equation}
%==============================
where $N=Z/v$ is the total number of counterions and the Hamiltonian
$\CALH=\CALT+\CALV$ splits into kinetic and potential degrees of
freedom.  In the classical description employed here the kinetic part
$\CALT$ will contribute the usual factor $\lambda^{-3N}$ to the
partition function, where $\lambda$ is the thermal de Broglie
wavelength.  The potential energy can be expressed as
%==============================
\begin{eqnarray}
  \CALV
  & = &
  -N \sum_i \; \frac{\ell}{|\VECr_i|}
  + \frac{1}{2} \sum_{i\ne j} \; \frac{\ell}{|\VECr_i-\VECr_j|} \ .
 \label{eq:ap4}
\end{eqnarray}
After rescaling all length by $\ell$, i.e. introducing $\VECx :=
\VECr/\ell$, the total partition function can be written as
\begin{eqnarray}
  \CALZ & = &
  \frac{1}{N!}\left(\frac{\ell}{\lambda}\right)^{3N}
  \int_{x_0}^{x_0/\phi^{1/3}} \prod_k \; \infd^3 x_k
  \\
  & & \exp\bigg\{
  -N \sum_i \; \frac{1}{|\VECx_i|}
  +\frac{1}{2} \sum_{i\ne j} \; \frac{1}{|\VECx_i-\VECx_j|}\bigg\} \ .
  \nonumber
\end{eqnarray}
In this form it becomes evident that appropriately scaled thermal
observables like the integrated charge density (measured in units of
$\ell^{-3}$) or the pressure (measured in units of $k_\rB T\ell^{-3}$)
are invariant under system changes which leave the number of
counterions $N$, the rescaled colloid size $x_0=r_0/\ell$, and the
volume fraction $\phi$ fixed.

Poisson-Boltzmann theory shows the same invariance property, as does
the approximate density functional theory we are proposing in this
paper.

%%%%%%%%%%%%%%%%%%%%%%%%%%%%%%%%%%%%%%%%%%%%%%%%%%%%%%%%%%%%%%%%%%%%%

%%%%%%%%%%%%%%%%%%%%%%%%%%%%%%%%%%%%%%%%%%%%%%%%%%%%%%%%%%%%%%%%%%%%%%%

\end{document}